\documentclass[12pt,a4paper]{iopart}
\usepackage[english]{babel}

\usepackage{iopams}
\usepackage{verbatim}
\usepackage{color}

\usepackage{graphicx}

\usepackage{color}

\usepackage[colorlinks,citecolor=blue,linkcolor=blue,urlcolor=blue]{hyperref}

\def\iotabar{\lower3pt\hbox{$\mathchar'26$}\mkern-7mu\iota}
\newcommand {\aplt}{\ {\raise-.5ex\hbox{$\buildrel<\over\sim$}}\ }
\newcommand{\dd}{\mbox{d}}

\newcommand{\spe}{{\sigma}}

\newcommand{\eq}[1]{(\ref{#1})}
\newcommand{\bun}{\hat{\mathbf{b}}}

\newcommand{\bv}{\mathbf{v}}

\newcommand{\bR}{\mathbf{R}}

\newcommand{\bJ}{\mathbf{J}}

\newcommand{\bB}{\mathbf{B}}

\newcommand{\dotcross}{ \raise 0.65ex\hbox{${\scriptstyle {{_{\displaystyle \cdot}}\atop\times}}$} }
\newcommand{\crossdot}{ \raise 0.5ex\hbox{${\scriptstyle {{_\times}\atop{\displaystyle \cdot}}}$} }

\newcommand{\sumsig}{ \raise -1.3ex\hbox{${{\displaystyle \sum}\atop{\scriptstyle \sigma}}$} }
\newcounter{appnumb}

\begin{document}

\title[Flow damping in stellarators close to quasisymmetry]
{Flow damping in stellarators close to quasisymmetry}
\author{Iv\'an Calvo$^{1}$}
\vspace{-0.2cm}
\eads{\mailto{ivan.calvo@ciemat.es}}
\vspace{-0.5cm}
\author{Felix I Parra$^{2,3}$}
\vspace{-0.2cm}
\eads{\mailto{felix.parradiaz@physics.ox.ac.uk}}
\vspace{-0.5cm}
\author{Jos\'e Luis Velasco$^{1}$}
\vspace{-0.2cm}
\eads{\mailto{joseluis.velasco@ciemat.es}}
\vspace{-0.5cm}
\author{J Arturo Alonso$^{1}$}
\vspace{-0.2cm}
\eads{\mailto{arturo.alonso@externos.ciemat.es}}

\vspace{1cm}

\address{$^1$Laboratorio Nacional de Fusi\'on, CIEMAT, 28040 Madrid, Spain}
\address{$^2$Rudolf Peierls Centre for Theoretical Physics, University of Oxford, Oxford, OX1 3NP, UK}
\address{$^3$Culham Centre for Fusion Energy, Abingdon, OX14 3DB, UK}


\vskip 1cm

{\large
\begin{center}
\today
\end{center}
}

\vspace{1cm}

\begin{abstract}
  Quasisymmetric stellarators are a type of optimized stellarators for
  which flows are undamped to lowest order in an expansion in the
  normalized Larmor radius. However, perfect quasisymmetry is
  impossible. Since large flows may be desirable as a means to reduce
  turbulent transport, it is important to know when a stellarator can
  be considered to be sufficiently close to quasisymmetry. The answer
  to this question depends strongly on the size of the spatial
  gradients of the deviation from quasisymmetry and on the
  collisionality regime. Recently, criteria for closeness to
  quasisymmetry have been derived in a variety of situations. In
  particular, the case of deviations with large gradients was solved
  in the $1/\nu$ regime. Denoting by $\alpha$ a parameter that gives
  the size of the deviation from quasisymmetry, it was proven that
  particle fluxes do not scale with $\alpha^{3/2}$, as typically
  claimed, but with $\alpha$. It was also shown that ripple wells are
  not the main cause of transport. This paper reviews
  those works and presents a new result in another collisionality
  regime, in which particles trapped in ripple wells are collisional
  and the rest are collisionless.
\end{abstract}

\maketitle

\section{Introduction}
\label{sec:Introduction}

Microturbulence is the main cause of transport in tokamaks and in the
outer region of stellarator plasmas~\cite{Helander2012}. Since flow
shear reduces microturbulent transport~\cite{Connor2004}, it is
desirable to understand under which conditions large flow shear can be
achieved. Even if not the only one, a reasonable route to large flow
shear consists of designing magnetic confinement devices that admit
large flows or, equivalently, that have weak flow damping.

Whereas a tokamak has inherently weak damping of flows in the toroidal
direction due to axisymmetry, this does not need to be the case for a
stellarator, even if its magnetic configuration has been
neoclassically optimized, i.e. even if it is such that collisionless
trajectories are confined. A stellarator optimized for neoclassical
transport is called omnigenous and its defining property is the
vanishing of the average of the radial magnetic drift along the lowest
order trapped orbits. The stellarator Wendelstein 7-X has been
designed to be omnigenous~\cite{nuhrenberg95}. Whereas omnigeneity is
a necessary condition for weak flow damping, it is not sufficient. It
has been proven~\cite{Helander08} that there exists a direction in
which flows are undamped if and only if the stellarator is
quasisymmetric. A magnetic field is
quasisymmetric~\cite{Boozer83,Nuehrenberg88} when, in Boozer
coordinates (see below), its magnitude depends only on a single
helicity; that is, on a single linear combination of the Boozer
poloidal and toroidal angles. In a quasisymmetric stellarator, the
guiding-center equations are equivalent to those of a tokamak to
lowest order. They possess a symmetry direction, and flows are
undamped along that direction.

  Technological issues place relevant bounds to how omnigeneous a real
  stellarator can be, but there is no theoretical reason to believe
  that omnigenous stellarators cannot be built. This is not the case
  for quasisymmetric magnetic fields. By means of an expansion in the
  inverse aspect ratio, Garren and Boozer proved~\cite{Garren1991}
  that quasisymmetry can be satisfied up to second order, but it is
  necessarily violated to third order in the expansion
  parameter. Then, it only makes sense to talk about stellarators
  close to quasisymmetry. It is actually possible to be close to
  quasisymmetry, as the existence of the HSX
  stellarator~\cite{Anderson1995} shows. An experimental study
    of flow damping in HSX can be found in \cite{Gerhardt2005}. Let
  $\bB = \bB_0 + \alpha \bB_1$ be the magnetic field, where $\bB_0$ is
  quasisymmetric, $\alpha \bB_1$ is the deviation from quasisymmetry
  and $0 \le \alpha \ll 1$ is a small parameter. The theoretical
  question is to understand when $\alpha$ is sufficiently small for
  the stellarator to be considered quasisymmetric in practice.

The systematic approach to this question was started in references
\cite{CalvoParraVelascoAlonso13} and \cite{CalvoParraAlonsoVelasco14}
by exploiting the equivalence of quasisymmetry and intrinsic
ambipolarity~\cite{Helander08}, i.e. the fact that
\begin{eqnarray}
\left\langle
\bJ_n\cdot\nabla\psi
\right\rangle_\psi = 0
\end{eqnarray}
for any density, temperature and radial electric field profiles
is equivalent to having no flow damping. Here, $\bJ_n$
  is the neoclassical electric current density and $\psi$ is a
  flux-label coordinate. In the previous equation $\langle \cdot
  \rangle_\psi$ denotes the flux surface average, defined as
\begin{eqnarray}
  \langle
  f
  \rangle_\psi
  =
  \frac{1}{V'}\int_0^{2\pi}\int_0^{2\pi} \sqrt{g} f(\psi,\Theta,\zeta)\,\dd\Theta\dd\zeta,
\end{eqnarray}
where $\Theta$ is a poloidal angle, $\zeta$ a toroidal angle,
$\sqrt{g}$ is the square root of the metric determinant, $V(\psi)$ is
the plasma volume enclosed by the surface labeled by $\psi$, and its
derivative is given by
\begin{eqnarray}
V'(\psi)
=
\int_0^{2\pi}\int_0^{2\pi} \sqrt{g}\,\dd\Theta\dd\zeta.
\end{eqnarray}

In reference \cite{Helander08} it is shown that in a stellarator far
from quasisymmetry flow damping is given precisely by $\left\langle
  \bJ_n\cdot\nabla\psi \right\rangle_\psi$. Thus, this quantity
vanishes identically for a quasisymmetric magnetic field, i.e. for
$\alpha = 0$. Therefore, it is the scaling of $\left\langle
  \bJ_n\cdot\nabla\psi \right\rangle_\psi$ with $\alpha$ that needs to
be determined in order to derive criteria for closeness to
quasisymmetry. How the determination of the scaling allows to obtain
the criterion is explained in Section
\ref{sec:formulationOfTheProblem}.

The calculation of the scalings, and the scalings themselves, have
revealed to be quite different depending on the size of the gradients
of the perturbation, and so will be the criteria derived from
them. Let us give a summary of some of the results contained in
references \cite{CalvoParraVelascoAlonso13} and
\cite{CalvoParraAlonsoVelasco14}. From now on, we assume that
$(\psi,\Theta,\zeta)$ are Boozer coordinates~\cite{Boozer81}. They
exist as long as $\bJ\cdot\nabla\psi = 0$, where $\bJ$ is the total
electric current density, and are defined by requiring that the
magnetic field can be simultaneously written as
\begin{eqnarray}
\fl
\bB = -\tilde\eta\nabla\psi + \frac{I_t(\psi)}{2\pi}\nabla\Theta +
\frac{I_p(\psi)}{2\pi}\nabla\zeta
\end{eqnarray}
and as
\begin{eqnarray}
\fl
\bB = \frac{\Psi'_p(\psi)}{2\pi}\nabla\zeta\times\nabla\psi +
\frac{\Psi'_t(\psi)}{2\pi}\nabla\psi\times\nabla\Theta.
\end{eqnarray}
Here, $I_t$ and $I_p$ are flux functions, $\Psi_t$ is the toroidal
flux, $\Psi_p$ is the poloidal flux, $\tilde\eta(\psi,\Theta,\zeta)$
is a singly-valued function and primes denote differentiation with
respect to $\psi$. Two properties of Boozer coordinates will be
relevant for us. The first one is that $\sqrt{g}$ takes the form
\begin{equation}\label{eq:sqrtgBoozer}
\sqrt{g} = \frac{V' \langle B^2 \rangle_\psi}{4\pi^2 B^2}.
\end{equation}
The second one is that if $\bB$ is quasisymmetric, then its magnitude
$B$ depends on a single linear combination, or helicity, of the Boozer
angles.

With the above notation, $B_0$ depends only on one helicity,
$B_0\equiv B_0(\psi, M\Theta - N\zeta)$, and we can take
$B_1(\psi,\Theta,\zeta)$ such that it does not contain the helicity
$M\Theta-N\zeta$. Then, we look at $|\partial_\Theta
B_1|/|\partial_\Theta B_0|$ and $|\partial_\zeta B_1|/|\partial_\zeta
B_0|$. As shown in reference \cite{CalvoParraVelascoAlonso13}, if
\begin{eqnarray}\label{eq:conditionSmallGradient}
  \frac{|\alpha\partial_\Theta
    B_1|}{|\partial_\Theta B_0|}\sim \alpha,
\nonumber
\\[5pt]
\frac{|\alpha\partial_\zeta
    B_1|}{|\partial_\zeta B_0|}\sim \alpha,
\end{eqnarray}
then 
\begin{eqnarray}\label{eq:quadraticScaling}
\fl
\left\langle
\bJ_n\cdot\nabla\psi
\right\rangle_\psi \sim \alpha^2 k
\end{eqnarray}
for any value of the collisionality (the quadratic scaling was
obtained in \cite{Simakov2009} for high collisionality). The
scaling of the function $k$ with the collision frequency depends on the
collisionality regime. If, on the contrary,
\begin{eqnarray}\label{eq:largeGradients}
  \frac{|\alpha\partial_\Theta
    B_1|}{|\partial_\Theta B_0|}\sim 1,
\\[5pt]
\frac{|\alpha\partial_\zeta
    B_1|}{|\partial_\zeta B_0|}\sim 1,
\end{eqnarray}
without further assumptions, then
\begin{eqnarray}
  \fl
  \left\langle
    \bJ_n\cdot\nabla\psi
  \right\rangle_\psi = O(\alpha^0)
\end{eqnarray}
and the quasisymmetric properties of $\bB_0$ have been completely
destroyed by the perturbation. However, as explained in
\cite{CalvoParraAlonsoVelasco14}, if perturbations with large
gradients are present and the extra assumption
\begin{equation}\label{eq:AssumptionMagDrift}
v_{M,\psi} - v_{M,\psi}^{(0)} \ll v_{M,\psi}^{(0)}
\end{equation}
is made, where $v_{M,\psi}^{(0)}$ is the radial magnetic drift
corresponding to $\bB_0$, then a scaling less favorable than
$\alpha^2$ but more favorable than $\alpha^0$ is obtained. Hence, when
designing a stellarator close to quasisymmetry large helicity
perturbations should be avoided; but if this is not possible, and it
probably is not, condition \eq{eq:AssumptionMagDrift} should,
  in principle\footnote{In reference \cite{Parra2014} it is
      proven that the extra assumption \eq{eq:AssumptionMagDrift} can
      actually be relaxed in the $1/\nu$ regime. However, we do not
      have a similar proof for other collisionality regimes.}, be a
  design goal.

Denote by $L_0$ the typical variation length of $B_0$, $L_0^{-1} \sim
|\nabla\ln B_0|$. In particular, the wells of $B_0$ along a magnetic
field line have size $L_0$.  It is easy to realize that when the
perturbation $\alpha\bB_1$ has strong gradients, $\bB = \bB_0 + \alpha
\bB_1$ can have, in addition, ripple wells of size $L_1\sim |\nabla\ln
B_1|^{-1}$ (see figure
\ref{fig:PerturbedQSfieldLine}). Sometimes~\cite{Beidler2011},
  it has been mistakenly thought that the original results of
  \cite{Ho1987} apply to stellarators close to quasisymmetry. This has
  led to the assumption that ripple wells are the main cause of
  transport, and that in the so-called $1/\nu$ regime particle and
energy fluxes scale with $\alpha^{3/2}$. In
\cite{CalvoParraAlonsoVelasco14}, it was shown that this is not
correct. A rigorous calculation was carried out assuming
\eq{eq:largeGradients} and \eq{eq:AssumptionMagDrift} and the result
was found to be, in the $1/\nu$ regime,
\begin{equation}\label{eq:linearScaling}
  \left\langle
    \bJ_n\cdot\nabla\psi
  \right\rangle_\psi\sim
  \frac{\alpha \epsilon_i^2 v_{ti}}{L_0\nu_{ii}} e n_i v_{ti} 
|\nabla\psi|_0.
\end{equation}
Furthermore, the calculation shows that ripple wells do not dominate
transport. Even if they contribute to the fluxes with the scaling
given on the right side of \eq{eq:linearScaling}, the same scaling is
produced by orbits trapped in the wells of size $L_0$; more
specifically, by the part of the orbit near the bounce
points. Here, $v_{ti}$ is the ion thermal speed, $n_i$ is the
  ion equilibrium density, $\nu_{ii}$ is the ion-ion collision
  frequency, $|\nabla\psi|_0$ is a characteristic value of
  $|\nabla\psi|$ on the flux surface, and $\epsilon_i := \rho_i/L_0$,
  where $\rho_i$ is the ion Larmor radius. Throughout the paper the
ion and electron temperatures are assumed to be comparable, and an
expansion in the square root of the ratio of the electron and ion
masses, $\sqrt{m_e/m_i}$, is employed.

The rest of the paper gives a brief summary of the calculations leading
to the above results and presents the criteria for closeness to
quasisymmetry implied by them. We also derive the scaling of the
fluxes with $\alpha$, and the corresponding criterion, in a new
collisionality regime. In \cite{CalvoParraAlonsoVelasco14} passing
particles, particles trapped in wells of size $L_0$, and particles
trapped in wells of size $L_1$ were assumed collisionless (this is
what is understood by $1/\nu$ regime). Here, we extend the computation
to the case when particles in ripple wells are collisional and the
rest are collisionless. Finally, we provide a treatment of the
collisional boundary layers more detailed than in our previous
works. The results of \cite{CalvoParraVelascoAlonso13} and
\cite{CalvoParraAlonsoVelasco14}, and those presented for the first
time here, also apply to tokamaks with ripple.

Before finishing this Introduction, it is pertinent to comment
  on the size of the plasma flows that we are dealing with. In this
  paper we always assume that the ion flow $V_i$ is subsonic,
  i.e. $V_i \ll v_{ti}$. The presence of sonic flows,
  $V_i\sim v_{ti}$, immediately implies
  quasisymmetry~\cite{Helander2007}. Hence, the absence of damping of
  subsonic flows is a necessary condition to achieve sonic
  flows. However, it is not sufficient. It has recently been
  proven~\cite{Sugama11} that even in quasisymmetric stellarators
  there are global obstructions that forbid strictly sonic flows, but
  they do not exclude regimes with $\epsilon_i v_{ti}\ll V_i \ll
  v_{ti}$. The actual flow size that can be reached in a stellarator
  that is not perfectly quasisymmetric will be the subject of further
  research.

\section{Formulation of the problem}
\label{sec:formulationOfTheProblem}

We use the subindex $\spe$ to denote different species and define
$\epsilon_\spe := \rho_\spe/L_0$, the Larmor radius over the
macroscopic scale. We also need to define the collisionality
$\nu_{*\spe\spe'} = \nu_{\spe\spe'}L_0/v_{t\spe}$, where
$\nu_{\spe\spe'}$ is the collision frequency of species $\spe$ with
$\spe'$. Here, $v_{t\spe}=\sqrt{T_\spe/m_\spe}$, $T_\spe$ and $m_\spe$
are the thermal speed, temperature and mass of species $\spe$. Denote
by $\nu_{*\spe}$ the largest of all collisionalities
$\nu_{*\spe\spe'}$ when $\spe'$ runs over species.  If $\epsilon_\spe
\ll \nu_{*\spe}$ for all species, one can eliminate the degree of
motion corresponding to the gyration of particles around the magnetic
field by expanding the fields and the kinetic equations in
$\epsilon_\spe\ll 1$ and averaging order by order in the gyrophase.

We employ phase-space coordinates $(\psi,\Theta,\zeta,v,\lambda,s)$,
where $v$ is the magnitude of the velocity, $\lambda =
B^{-1}v_\perp^2/v^2$ is the pitch-angle coordinate, $v_\perp$ is the
magnitude of the velocity component perpendicular to the magnetic
field, and $s = \pm 1$ is the sign of the parallel velocity $v_{||}$,
that can be expressed as
\begin{equation}
v_{||} = s v \sqrt{1-\lambda B}\,.
\end{equation}
Let us use the notation $F_{\spe} = F_{\spe 0} + F_{\spe 1} + \dots$,
$\varphi = \varphi_0 + \varphi_1 + \dots$ for the expansions of the
distribution function and the electrostatic potential, where $F_{\spe
  1}/F_{\spe 0} = O(\epsilon_\spe)$, $\varphi_1/\varphi_0 =
O(\epsilon_i)$, $e \varphi_0 /T_i = O(1)$ and $e$ is the proton
charge.  To lowest order, one obtains that the distribution function
is Maxwellian
\begin{eqnarray}
&& \fl F_{\spe 0}(\bR,u,\mu)=
n_{\spe}
\left(\frac{m_\spe}{2\pi T_{\spe}}\right)^{3/2}
\exp\left(-\frac{m_\spe(u^2/2 + \mu B)}{T_{\spe}}\right)
\end{eqnarray}
and that the density $n_\spe$, the temperature $T_\spe$ and
$\varphi_0$ are flux functions. The next order pieces $F_{\spe 1}$ and
$\varphi_1$ have both neoclassical and turbulent components, but only
the former, that varies in macroscopic length scales, matters for our
calculation. Thus, $F_{\spe 1}$ and $\varphi_1$ stand for the
neoclassical components of the corrections to $F_{\spe 0}$ and
$\varphi_0$. Denote by $G_{\spe 1} = F_{\spe 1} + (Z_\spe e \varphi_1
/ T_\spe) F_{\spe 0}$ the non-adiabatic component of the correction to
the distribution function. The equation that determines $G_{\spe 1}$
is called {\em the drift-kinetic equation}~\cite{Hazeltine73} and reads
\begin{eqnarray}\label{eq:DKE}
v_{||}\bun\cdot\nabla G_{\spe 1} + \Upsilon_\spe v_{\psi,\spe} F_{\spe 0}
= C_\spe^{\ell}[G_{1}].
\end{eqnarray}
Here,
\begin{eqnarray}
\fl
\Upsilon_\spe := \frac{Z_\spe e }{T_\spe}\partial_\psi\varphi_0 +
\frac{1}{n_\spe}\partial_\psi n_\spe
\nonumber\\[5pt]
\fl\hspace{1cm}
+
\left(\frac{m_\spe v^2}{2T_\spe}-\frac{3}{2}\right)
\frac{1}{T_\spe}\partial_\psi T_\spe,
\end{eqnarray}
$v_{\psi,\spe} =
\bv_{M,\spe}\cdot\nabla\psi$ is the radial magnetic drift,
\begin{eqnarray}
  \bv_{M,\spe} = \frac{v^2(1-\lambda B)}{\Omega_\spe}\bun
\times(\bun\cdot\nabla\bun) + 
\frac{v^2\lambda}{2\Omega_\spe}\bun\times\nabla B,
\end{eqnarray}
$\Omega_\spe = Z_\spe e B /(m_\spe c)$ is the gyrofrequency of species
$\spe$, $Z_\spe e$ is the electric charge, $c$ is the speed of light
and $C_\spe^{\ell}[G_{1}]$ is the linearized collision operator, whose
explicit expression is not needed (see, for example,
\cite{helander02bk}).

The well-known neoclassical expression for the
flux-surface averaged radial electric current is
\begin{equation}\label{eq:neoclassicalJ}
\fl\left\langle
    \bJ_n\cdot\nabla\psi
  \right\rangle_\psi
  =
  \left\langle
  \sum_\spe\sum_{s=-1}^1 Z_\spe e\int_0^\infty
  \dd v
  \int_0^{B^{-1}}\dd\lambda\,
\frac{\pi v^2 B}{\sqrt{1-\lambda B}}v_{\psi,\spe}  G_{\spe 1}
  \right\rangle_\psi.
\end{equation}
From here on, and for the sake of simplicity, we only deal with the
ion-drift kinetic equation. We note that if the ion and electron
temperatures are comparable and just lowest-order terms in an
expansion in $\sqrt{m_e/m_i}$ are kept, then only the ion particle
flux needs to be taken into account in \eq{eq:neoclassicalJ} and
ion-electron collisions can be neglected.

Consider an expansion in $\epsilon_i$ of the total
  flux-surface averaged radial electric current, $\langle
  \bJ\cdot\nabla\psi\rangle_\psi$ (which, of course, vanishes due to
  quasineutrality). In a generic stellarator the right-hand side of
  \eq{eq:neoclassicalJ} dominates because it scales with
  $\epsilon_i^2$, whereas turbulent and higher-order neoclassical
  contributions to the total radial current are $O(\epsilon_i^3)$, as
  explained in detail in reference
  \cite{CalvoParraVelascoAlonso13}. These $O(\epsilon_i^3)$
  contributions include higher-order flow damping terms and the
  polarization current.  Schematically,
\begin{eqnarray}\label{eq:schem1}
\left\langle
    \bJ\cdot\nabla\psi
  \right\rangle_\psi
=
(\epsilon_i^2 A + \epsilon_i^3 C) e n_i v_{ti}|\nabla\psi|_0 + \dots
\end{eqnarray}
For perfectly quasisymmetric stellarators, $A$ is identically
zero. Then, for stellarators close to quasisymmetry, it is expected
that $A \sim \tilde{A} \alpha^q \nu_{* i}^r$, with $q>0$ and
$\tilde{A}=O(1)$. Then,
\begin{eqnarray}\label{eq:schem2}
  \left\langle
    \bJ\cdot\nabla\psi
  \right\rangle_\psi
  =
  \left(\epsilon_i\frac{V_i}{v_{ti}} \tilde{A} \alpha^q \nu_{* i}^r + 
\epsilon_i^3 C\right) e n_i v_{ti}|\nabla\psi|_0.
\end{eqnarray}
Note that we have replaced one instance of $\epsilon_i$ by
$V_i/v_{ti}$ in the first term of \eq{eq:schem1}. We have done this
just to point out that, in general, this is the actual scaling of that
term; this comes from the fact that $G_{\spe 1}$ depends linearly on
$\varphi_0'$, and the latter sets the magnitude of the flow. The two
terms in equation \eq{eq:schem2} are comparable when $V_i\sim
\epsilon_i^2 \alpha^{-q}\nu_{*i}^{-r} v_{ti}$. The stellarator
exhibits a quasisymmetric behavior if flows of size $V_i > \epsilon_i
v_{ti}$ are reached, which gives the criterion
\begin{equation}
  \alpha < (\epsilon_i \nu_{* i}^{-r})^{1/q}.
\end{equation}
Without further information, we have assumed $C=O(1)$ above.  Our
task consists of finding the powers $q$ and $r$, that depend on the
geometry of the deviation from quasisymmetry and on the collisionality
regime.

\section{Criteria for closeness to quasisymmetry}
\label{sec:rotationCriteria}

In this section we compute the scaling with $\alpha$ (and with
$\nu_{*i}$ when the result depends on the collisionality regime) of
the right side of \eq{eq:neoclassicalJ}. As advanced in the
Introduction, this depends strongly on the size of the gradient of
$B_1$, the magnitude of the perturbation to the quasisymmetric
configuration $\bB_0$.

\subsection{Perturbations with small gradients}
\label{sec:smallGradients}

The calculation when \eq{eq:conditionSmallGradient} is satisfied was
carried out in \cite{CalvoParraVelascoAlonso13} and reviewed in
\cite{CalvoParraAlonsoVelasco14}. Hence, we do not repeat it here. We
simply recall that it consists of Taylor expanding the right side of
\eq{eq:neoclassicalJ} written in Boozer coordinates. The result is
given in equation \eq{eq:quadraticScaling} and therefore the
stellarator can be considered quasisymmetric if
\begin{equation}\label{eq:criterionSmallGradients}
\alpha < \epsilon_i^{1/2},
\end{equation}
for a generic value $\nu_{*i}\sim 1$. In the regime $1/\nu$ the
function $k$ appearing in \eq{eq:quadraticScaling} scales with
$\nu_{*i}^{-1}$ and the criterion \eq{eq:criterionSmallGradients}
can be more precisely written as
\begin{equation}\label{eq:criterionSmallGradients1overNu}
  \alpha < \sqrt{\nu_{*i}\epsilon_i}.
\end{equation}

In the next subsection we show that when $B_1$ has large gradients the
calculation is more complicated and that the scaling with $\alpha$ is
not quadratic, but more unfavorable.

\subsection{Perturbations with large gradients}
\label{sec:largeGradients}

The straightforward approach of Taylor expanding the
  drift-kinetic equation \eq{eq:DKE} fails when the perturbation has
  large gradients in the sense of \eq{eq:largeGradients}. For example,
  it is clear that the parallel streaming operator $s v
  \sqrt{1-\lambda B}\, \bun\cdot\nabla$ cannot be expanded at points
  where $\nabla B_1\sim \alpha^{-1}L_0^{-1}B_0$.

As mentioned in the Introduction, if \eq{eq:largeGradients} is
satisfied and no additional condition is imposed, the perturbation to
the source term of the drift-kinetic equation \eq{eq:DKE}, $
v_{M,\psi} - v_{M,\psi}^{(0)}$, is $O(1)$ and the stellarator cannot
be viewed as a perturbation of a quasisymmetric one. Hence, we
assume $v_{M,\psi} - v_{M,\psi}^{(0)} \ll v_{M,\psi}^{(0)}$.

In reference \cite{CalvoParraAlonsoVelasco14} a detailed calculation
of the scaling has been given in the $1/\nu$ regime, i.e. when passing
particles, particles trapped in larges wells, and particles trapped in
ripple wells are collisionless. Here, we will extend those results to
the case when passing particles and particles trapped in larges wells
are collisionless, but particles trapped in ripple wells are
collisional.

\begin{figure}
\includegraphics[width=\columnwidth]{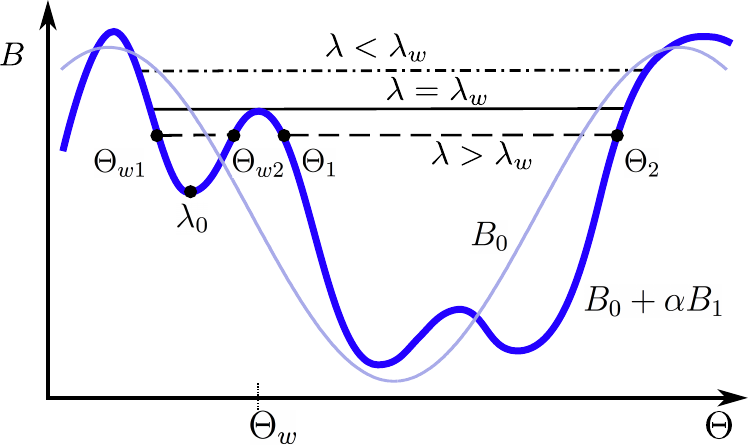}
\caption{Magnitude of a quasisymmetric magnetic field along a field
  line segment (thin curve), and magnitude of the perturbed field
  (thick curve).}
\label{fig:PerturbedQSfieldLine}
\end{figure}

It will be useful to employ $\{\psi,\chi,\Theta\}$ as spatial
coordinates, with $\chi = \Theta - \iotabar\zeta$. The coordinate
$\chi$ acts as a magnetic field line label and $\Theta$ as the
coordinate that gives the position along the magnetic field line. We
make $\epsilon_i \ll \nu_{*i} \ll 1$, that defines what is usually
understood by {\em low collisionality}.  The drift-kinetic equation
and its solutions can then be expanded in $\nu_{*i}$. Taking
\begin{equation}
G_i = G_i^{[-1]} + G_i^{[0]} + O(\nu_{*i}\epsilon_iF_{i0}),
\end{equation}
with $G_i^{[n]}\sim \nu_{*i}^n\epsilon_iF_{i0}$, one finds that
\begin{equation}
\bun\cdot\nabla  G_i^{[-1]} = 0.
\end{equation}
For passing particles this means that $G_i^{[-1]}$ is a flux function,
whereas for trapped particles it implies that $G_i^{[-1]}$ does not
depend on $\Theta$ and on the sign of the parallel velocity,
$s$. Hence, one can write
\begin{equation}
  G_i^{[-1]}(\psi,\chi,v,\lambda,s) =
  g_i(\psi,v,\lambda,s) 
+ \partial_\chi h_i (\psi,\chi,v,\lambda),
\end{equation}
with $h_i \equiv 0$ in the passing region. The function $G_i^{[-1]}$
is determined by going to next order in the $\nu_{*i}$ expansion,
\begin{eqnarray}\label{eq:DKEnexttolowestorder}
v_{||}\bun\cdot\nabla G_i^{[0]} + \Upsilon_i v_{\psi,i} F_{i 0}
= C_{ii}^{\ell}[G_i^{[-1]}].
\end{eqnarray}

For trapped trajectories, we multiply \eq{eq:DKEnexttolowestorder} by
$(v_{||}\bun\cdot\nabla\Theta)^{-1}$ and integrate over the orbit,
finding the constraint
\begin{eqnarray}\label{eq:ConstraintEq1overNuRegime}
\oint\Upsilon_i\frac{v_{\psi,i}}{v_{||}\bun\cdot\nabla\Theta}
\, F_{i 0}\dd\Theta
= \oint\frac{C^\ell_{ii}[G_i^{[-1]}]}{v_{||}\bun\cdot\nabla\Theta}\dd\Theta.
\end{eqnarray}

\begin{figure}
\includegraphics[width=\columnwidth]{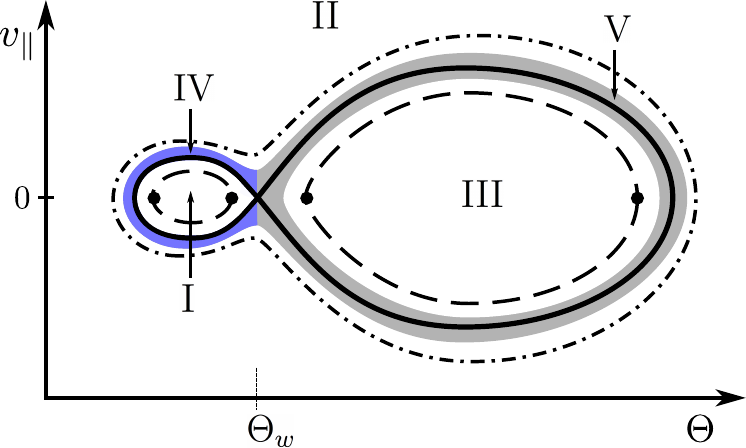}
\caption{The different regions into which phase space is divided to
  solve the drift-kinetic equation.}
\label{fig:Regions_PhaseSpace}
\end{figure}

An entropy production argument shows (see reference
\cite{CalvoParraAlonsoVelasco14}) that $g_\spe \equiv 0$ up to terms
$O(\alpha)$. We will see below that terms $O(\alpha)$ are negligible,
so that for our purposes $G_i^{[-1]}$ is zero in the passing region of
velocity space and
\begin{equation}
\int_0^{2\pi} G_i^{[-1]} \dd\chi = 0
\end{equation}
in the trapped region. Then, we focus on
\eq{eq:ConstraintEq1overNuRegime} to find the scaling of $G_i^{[-1]}$
for trapped particles. As discussed in the Introduction, when $B_1$
has large gradients, small wells of size $L_1\sim \alpha L_0$ are
typically created. Wells of size $L_0$ and ripple wells of size $L_1$
have to be treated separately. More specifically, the calculation is
arranged by dividing the trapped part of phase-space into the regions
depicted in figure \ref{fig:Regions_PhaseSpace}. Region I corresponds
to a ripple well of size $L_1$, and Regions II and III correspond to a
well of size $L_0$ already present in $B_0$.  We will give the
pertinent form of the drift-kinetic equation in each region, derive
the scaling of the distribution function, and then will investigate
the matching conditions. We will see that Regions IV and V are
collisional boundary layers that develop to heal discontinuities at
the interfaces among the first three regions.

 We start with particles trapped in  wells of size $L_0$. The key is
the computation of the scaling of the radial magnetic drift integrated
over the trajectory. The rigorous calculation is rather complicated
and is presented in detail in
\cite{CalvoParraAlonsoVelasco14}. Here, we only state the result,
\begin{eqnarray}\label{eq:ScalingIntMagDrift}
  \oint \frac{v_{\psi,i}}{v_{||}\bun\cdot\nabla\Theta}
  \dd\Theta
  \sim \alpha^{1/2}\psi.
\end{eqnarray}
Employing \eq{eq:ScalingIntMagDrift}, and the fact
  that in Regions II and III $\partial_\lambda \sim B_0$, it is easy
  to see that \eq{eq:ConstraintEq1overNuRegime} implies
\begin{eqnarray}\label{eq:SizeGRegionII}
 G_i^{\mathrm{II}} \sim \alpha^{1/2}\nu_{*i}^{-1}\epsilon_i F_{i0},
\end{eqnarray}
\begin{eqnarray}
G_i^{\mathrm{III}} \sim \alpha^{1/2} \nu_{*i}^{-1}\epsilon_i F_{i0}
\end{eqnarray}
and also
\begin{eqnarray}
\partial_\lambda G_i^{\mathrm{II}} \sim \alpha^{1/2} \nu_{*i}^{-1}\epsilon_i 
B_0 F_{i0},
\end{eqnarray}
\begin{eqnarray}
  \partial_\lambda G_i^{\mathrm{III}} \sim \alpha^{1/2}
 \nu_{*i}^{-1}\epsilon_i B_0 F_{i0}.
\end{eqnarray}
Observe that the size of these pieces of the distribution function
justifies neglecting terms $O(\alpha)$ of the distribution function in
the passing region, as advanced above.

Regions II and III give contributions to the 
neoclassical radial electric current
\begin{eqnarray}\label{eq:ScalingRegionII}
\left\langle
    \bJ_n\cdot\nabla\psi
  \right\rangle_\psi^{\mathrm{II}} \sim 
\frac{\alpha}{\nu_{*i}}\epsilon_i^2 e n_i v_{ti} |\nabla\psi|_0
\end{eqnarray}
and
\begin{eqnarray}\label{eq:ScalingRegionIII}
\left\langle
    \bJ_n\cdot\nabla\psi
  \right\rangle_\psi^{\mathrm{III}} \sim 
\frac{\alpha}{\nu_{*i}}\epsilon_i^2 e n_i v_{ti} |\nabla\psi|_0.
\end{eqnarray}

We turn to find the distribution function in the ripple well,
Region ${\mathrm{I}}$. We want to solve the regimes
\begin{equation}
\alpha^{-1/2}\nu_{*i} \ll 1,
\end{equation}
corresponding to the case in which particles trapped in ripple wells
are collisionless (the case solved in
\cite{CalvoParraAlonsoVelasco14}), and
\begin{equation}
\alpha^{-1/2}\nu_{*i} \gg 1,
\end{equation}
when those particles are collisional. For the moment, we take the
maximal ordering $\alpha^{-1/2}\nu_{*i}\sim 1$. Since particles in the
ripple wells are not necessarily collisionless, we cannot use
\eq{eq:ConstraintEq1overNuRegime}, but we have to resort to
\eq{eq:DKE}. However, the latter equation can be simplified to
\begin{eqnarray}\label{eq:DKEsimplified}
v_{||}\bun\cdot\nabla\Theta\partial_\Theta G_{i}^{\mathrm{I}}
-\frac{v_{||}\nu_\lambda}{v^2 B_{\mathrm{wmax}}^2}\partial_\lambda(v_{||}\partial_\lambda  G_{i}^{\mathrm{I}}) =
-\Upsilon_i v_{\psi,i} F_{i 0},
\end{eqnarray}
where $\nu_\lambda(v)$ is the pitch-angle scattering frequency. Here,
we have approximated $B \approx B_{\mathrm{wmax}} $ and
$\lambda\approx \lambda_w = B_{\mathrm{wmax}}^{-1}$, with $\lambda_w$
defined in figure \ref{fig:PerturbedQSfieldLine} along with other
quantities that will be employed below. In order to get the above
simplified equation, one needs the relations
\begin{eqnarray}
  \fl
  B_0^{-1}\partial_\lambda G_{i}^{\mathrm{I}} \gg
  v_{ti}\partial_v G_{i}^{\mathrm{I}} \sim v_{ti}\partial_v G_{i}^{\mathrm{II}}
  \nonumber\\[5pt]
  \fl\hspace{1cm}  \sim v_{ti}\partial_v G_{i}^{\mathrm{III}}
\sim B_0^{-1}\partial_\lambda G_{i}^{\mathrm{II}} \sim B_0^{-1}
\partial_\lambda G_{i}^{\mathrm{III}},
\end{eqnarray}
obtained by using $v_{||}\sim \alpha^{1/2}v_{ti}$,
$\bun\cdot\nabla\sim (\alpha L_0)^{-1}$ and $\partial_\lambda \sim
B_0/\alpha$.

If $\alpha^{-1/2}\nu_{*i}\ll 1$, then to lowest order
$G_{i}^{\mathrm{I}}$ does not depend on $\Theta$ and it is determined
by integrating \eq{eq:DKEsimplified} over the orbit,
\begin{eqnarray}\label{eq:DKEsimplifiedIntegrated}
\int_{\Theta_{w1}}^{\Theta_{w2}}\frac{\dd\Theta}{v_{||}\bun\cdot\nabla\Theta}
  \frac{v_{||}\nu_\lambda}{v^2 B_{\mathrm{wmax}}^2}
\partial_\lambda(v_{||}\partial_\lambda  G_{i}^{\mathrm{I}}) =
 \int_{\Theta_{w1}}^{\Theta_{w2}}\frac{\dd\Theta}{v_{||}\bun\cdot\nabla\Theta}
 \Upsilon_i v_{\psi,i} F_{i 0}.
\end{eqnarray}
Imposing a regularity condition at the bottom of the well
$\lambda_0$, we can find an explicit
expression for $\partial_\lambda G_{i}^{\mathrm{I}}$ in the
well. Using again that the size of the well in $\lambda$ is
$O(\alpha)$, we get
\begin{equation}
\partial_\lambda G_{i}^{\mathrm{I}}\sim \nu_{*i}^{-1}\epsilon_i F_{i0}.
\end{equation}

If $\alpha^{-1/2}\nu_{*i}\gg 1$, one can drop the parallel streaming
term in \eq{eq:DKEsimplified} to obtain
\begin{eqnarray}\label{eq:DKEsimplifiedCollisionalWell}
\frac{v_{||}\nu_\lambda}{v^2 B_{\mathrm{wmax}}^2}\partial_\lambda(v_{||}\partial_\lambda  G_{i}^{\mathrm{I}}) = \Upsilon_i v_{\psi,i} F_{i 0}.
\end{eqnarray}
This also yields $\partial_\lambda G_{i}^{\mathrm{I}}\sim
\nu_{*i}^{-1}\epsilon_i F_{i0}$. In this case $\partial_\lambda
G_{i}^{\mathrm{I}}$ can depend on $\Theta$.

Hence, the size of $\partial_\lambda G_{i}^{\mathrm{I}}$ is the same
for both regimes, $\alpha^{-1/2}\nu_{*i}\ll 1$ and
$\alpha^{-1/2}\nu_{*i}\gg 1$. Whereas we have been able to determine
the size of $\partial_\lambda G_{i}^{\mathrm{I}}$ by using only the
drift-kinetic equation in the ripple well, the size of
$G_i^{\mathrm{I}}$ is still unknown. It is determined by matching with
$G_i^{\mathrm{II}}$ and $G_i^{\mathrm{III}}$, and found to be,
therefore,
\begin{eqnarray}\label{eq:SizeGRegionI}
 G_i^{\mathrm{I}} \sim \alpha^{1/2}\nu_{*i}^{-1}\epsilon_i F_{i0}.
\end{eqnarray}
Assuming a number of ripple wells per magnetic field line
$O(\alpha^{-1})$, and a number of magnetic field lines with ripple
wells $O(\alpha^{-1})$, we find the contribution of Region I to the
flux-surface averaged radial current,
\begin{equation}\label{eq:ScalingRegionI}
\left\langle
    \bJ_n\cdot\nabla\psi
  \right\rangle_\psi^{\mathrm{I}}
\sim 
\frac{\alpha}{\nu_{*i}}\epsilon_i^2 e n_i v_{ti}|\nabla\psi|_0.
\end{equation}

In deriving the scalings \eq{eq:ScalingRegionII},
\eq{eq:ScalingRegionIII} and \eq{eq:ScalingRegionI}, we have skipped
the proof that the matching among the three regions can be done
consistently. Actually, it cannot be done without Regions IV and
V. One can show that the values of $\partial_\lambda G_i^{\mathrm{I}}$
and $\partial_\lambda G_i^{\mathrm{II}}$ do not match exactly, and the
same happens with $\partial_\lambda G_i^{\mathrm{II}}$ and
$\partial_\lambda G_i^{\mathrm{III}}$. The correct matching is
provided by the emergence of collisional layers at the interfaces
between I and II (denoted by Region IV) and II and III (denoted by
Region $V$). A careful treatment of the collisional layers, Regions IV
and V, gives
\begin{equation}
\left\langle
    \bJ_n\cdot\nabla\psi
  \right\rangle_\psi^{\mathrm{IV}}\sim 
\alpha\epsilon_i^2 e n_i v_{ti}|\nabla\psi|_0
\end{equation}
and
\begin{equation}
\left\langle
    \bJ_n\cdot\nabla\psi
  \right\rangle_\psi^{\mathrm{V}}\sim 
\alpha^{1/2}\epsilon_i^2 e n_i v_{ti}|\nabla\psi|_0,
\end{equation}
where the assumption that the number of ripple wells is
$O(\alpha^{-2})$ has been employed again. We postpone
  the details on the boundary layers to Section
  \ref{eq:CollBoundaryLayer}. In particular, the estimations of Section
  \ref{eq:CollBoundaryLayer} show that the
  discontinuities in the absence of the collisional boundary layers
  are small enough that the estimates \eq{eq:ScalingRegionII},
  \eq{eq:ScalingRegionIII} and \eq{eq:ScalingRegionI} are correct.

It is clear that the contribution of Region IV is always
negligible. When $\alpha^{-1/2}\nu_{*i}\ll 1$, Regions I, II and III
contribute the same and the contribution of Region V is negligible,
giving
\begin{equation}\label{eq:JcollisionlessRippleWells}
  \left\langle
    \bJ_n\cdot\nabla\psi
  \right\rangle_\psi\sim
\frac{\alpha}{\nu_{*i}} \epsilon_i^2  e n_i v_{ti}|\nabla\psi|_0.
\end{equation}
However, when $\alpha^{-1/2}\nu_{*i}\gg 1$, the contribution of Region
V dominates and one obtains
\begin{equation}\label{eq:JcollisionlessRippleWells}
  \left\langle
    \bJ_n\cdot\nabla\psi
  \right\rangle_\psi\sim
\frac{\alpha^{1/2} \epsilon_i^2 v_{ti}}{L_0} e n_i v_{ti}|\nabla\psi|_0.
\end{equation}

Finally, we write the criteria to assess closeness to quasisymmetry
inferred from the above results. Namely, if the perturbation has large
gradients and $\alpha^{-1/2}\nu_{*i}\ll 1$, then the criterion is
\begin{equation}
\alpha < \nu_{*i}\epsilon_i.
\end{equation}
If the perturbation has large helicities and $\alpha^{-1/2}\nu_{*i}\gg
1$, then the criterion is
\begin{equation}
\alpha < \epsilon_i^2.
\end{equation}
One can compare these criteria with
  \eq{eq:criterionSmallGradients1overNu} and confirm that, for the
  same value of $\alpha$, large helicity perturbations degrade more
  efficiently the quasisymmetric properties of the stellarator than
  small helicity ones, as expected.

\section{The collisional boundary layers}
\label{eq:CollBoundaryLayer}

Recall that the different regions into which we split phase space to
compute the distribution function are shown in figure
\ref{fig:Regions_PhaseSpace}. We have explained how to determine the
distribution function in Regions I, II, and III, and here we focus on
the collisional boundary layers, Regions IV and V. In fact, as already
pointed out, the correct treatment of the collisional boundary layers
is also needed to ensure that the results on Regions I, II, and III
are consistent. Below we show how to find two functions $\delta
G_i^{\mathrm{IV}}$ and $\delta G_i^{\mathrm{V}}$ such that
$G_i^{\mathrm{II}} + \delta G_i^{\mathrm{IV}}$ smoothly matches
$G_i^{\mathrm{I}}$, and $G_i^{\mathrm{II}} + \delta G_i^{\mathrm{V}}$
smoothly matches $G_i^{\mathrm{III}} + \delta
G_i^{\mathrm{V}}$. This section is based on the
  rigorous discussion given in \cite{Parra2014}. In this reference,
  the boundary layers are calculated for a stellarator close to
  omnigeneity, proving that their effect on transport is
  negligible. As advanced in subsection \ref{sec:largeGradients}, the
  boundary layers are more relevant in stellarators close to
  quasisymmetry because they can dominate flow
  damping in the regime in which particles trapped in ripple wells are
  collisional and the rest are collisionless.

Let us start by Region IV and recall the notation used in figure
\ref{fig:PerturbedQSfieldLine}. We define a function $\delta
G_i^{\mathrm{IV}}$ supported in $\lambda\in [\lambda_w -
K\delta\lambda_{\mathrm{IV}},\lambda_w]$ and
$\Theta\in[\Theta_{1}(\lambda),\Theta_w]$. Here,
$\delta\lambda_{\mathrm{IV}}$ is the characteristic width of the layer
and $K\gg 1$. The equation for $\delta G_i^{\mathrm{IV}}$ is
\begin{eqnarray}\label{eq:DKECollisionalLayerSmall}
v_{||}\bun\cdot\nabla\Theta\partial_\Theta \delta G_{i}^{\mathrm{IV}}
-\frac{v_{||}\nu_\lambda}{v^2 B_{\mathrm{wmax}}^2}
\partial_\lambda(v_{||}\partial_\lambda  \delta G_{i}^{\mathrm{IV}}) = 0,
\end{eqnarray}
which has to be solved with the boundary conditions
$\partial_\lambda\delta G_{i}^{\mathrm{IV}} = \partial_\lambda
G_{i}^{\mathrm{I}} - \partial_\lambda G_{i}^{\mathrm{II}}$ at
$\lambda = \lambda_w$, $\delta G_{i}^{\mathrm{IV}} = 0$ at $\lambda =
\lambda_w - K\delta\lambda_{\mathrm{IV}}$, $\delta G_{i}^{\mathrm{IV}}
= \delta G_{i}^{\mathrm{V}}$ at $\Theta = \Theta_w$, the interface
between the two collisional boundary layers, and $\delta
G_{i}^{\mathrm{IV}}(s=1) = \delta G_{i}^{\mathrm{IV}}(s=-1)$ at $\Theta_1$.

If both terms in \eq{eq:DKECollisionalLayerSmall} have to be of the same
order, and in the layer $v_{||}\sim\alpha^{1/2} v_{ti}$ and
$\bun\cdot\nabla \sim \alpha^{-1}L_0^{-1}$, one infers
\begin{equation}
\delta\lambda_{\mathrm{IV}}
\sim
\alpha^{3/4}\nu_{*i}^{1/2}B_0^{-1}.
\end{equation}
The boundary condition on $\partial_\lambda\delta G_{i}^{\mathrm{IV}}$
at $\lambda=\lambda_w$ gives
\begin{equation}
\partial_\lambda\delta G_{i}^{\mathrm{IV}}
\sim
B_0\nu_{*i}^{-1}\epsilon_i F_{i0},
\end{equation}
and therefore
\begin{equation}
\delta G_{i}^{\mathrm{IV}}
\sim
\alpha^{3/4}\nu_{*i}^{-1/2}\epsilon_i F_{i0}.
\end{equation}

We proceed to deal with the collisional layer denoted by Region
V. In this case, $\delta G_{i}^{\mathrm{V}}$ satisfies
\begin{eqnarray}\label{eq:DKECollisionalLayerLarge}
v_{||}\bun\cdot\nabla\Theta\partial_\Theta \delta G_{i}^{\mathrm{V}}
-\frac{v_{||}\nu_\lambda}{v^2 B_{\mathrm{wmax}}^2}
\partial_\lambda(v_{||}\partial_\lambda  \delta G_{i}^{\mathrm{V}}) = 0
\end{eqnarray}
and is defined on $\lambda\in [\lambda_w -
K\delta\lambda_{\mathrm{V}},\lambda_w + K\delta\lambda_{\mathrm{V}}]$,
$\Theta\in[\Theta_w,\Theta_{2}(\lambda)]$ for $\lambda < \lambda_w$,
and $\Theta\in [\Theta_{1}(\lambda), \Theta_{2}(\lambda)]$ for
$\lambda > \lambda_w$. The boundary conditions are $\partial_\lambda
\delta G_{i}^{\mathrm{V}}(\lambda_w^+) - \partial_\lambda \delta
G_{i}^{\mathrm{V}}(\lambda_w^-) = - (\partial_\lambda
G_{i}^{\mathrm{III}}(\lambda_w) - \partial_\lambda
G_{i}^{\mathrm{II}}(\lambda_w)) $ at $\lambda = \lambda_w$, $\delta
G_{i}^{\mathrm{V}} = 0$ at $\lambda_w - K\delta\lambda_{\mathrm{V}}$
and $\lambda_w + K\delta\lambda_{\mathrm{V}}$, $\delta
G_{i}^{\mathrm{IV}} = \delta G_{i}^{\mathrm{V}}$, for $\lambda <
\lambda_w$ at $\Theta = \Theta_w$, $\delta G_{i}^{\mathrm{V}}(s=1) =
\delta G_{i}^{\mathrm{V}}(s=-1)$ for $\lambda > \lambda_w$ at
$\Theta_1$, and $\delta G_{i}^{\mathrm{V}}(s=1) = \delta
G_{i}^{\mathrm{V}}(s=-1)$, for any $\lambda$, at $\Theta_2$.

Balancing the two terms in \eq{eq:DKECollisionalLayerLarge}, and
employing $v_{||}\sim v_{ti}$ and $\bun\cdot\nabla \sim L_0^{-1}$, one
gets
\begin{equation}
\delta\lambda_{\mathrm{V}}
\sim
\nu_{*i}^{1/2}B_0^{-1}.
\end{equation}
Estimating the size of $\delta G_{i}^{\mathrm{V}}$ is slightly more
involved. It can be shown~\cite{Parra2014} that phase space continuity
between Regions IV and V implies
\begin{equation}
\delta G_{i}^{\mathrm{V}}
\sim
\alpha^{3/2}\nu_{*i}^{-1/2}\epsilon_i F_{i0}.
\end{equation}

Finally, we emphasize that the size of both $\delta
G_{i}^{\mathrm{IV}}$ and $\delta G_{i}^{\mathrm{V}}$ is very small
compared to $G_{i}^{\mathrm{II}}$ and $G_{i}^{\mathrm{III}}$, given in
\eq{eq:SizeGRegionII}. This implies, in particular, that the
estimation \eq{eq:SizeGRegionI} is correct.

\section{Conclusions}
\label{eq:Conclusions}

In this paper we have first reviewed the results of references
\cite{CalvoParraVelascoAlonso13} and
\cite{CalvoParraAlonsoVelasco14}. In those references, we started a systematic
approach to understand how much one can deviate
from perfect quasisymmetry without spoiling some of the properties
that make quasisymmetric stellarators interesting; in particular, the
absence of flow damping in the symmetry direction. We show that the
answer depends on the size of the helicity of the deviations from
quasisymmetry and, in general, on the collisionality regime. Formal
criteria are derived in a variety of situations, to assess whether a
stellarator can be considered quasisymmetric in practice. The survey
of \cite{CalvoParraVelascoAlonso13} and
\cite{CalvoParraAlonsoVelasco14} is completed by extending those results
to a new collisionality regime. All these results apply to tokamaks
with ripple as well.

\ack

This work has been carried out within the framework of the EUROfusion
Consortium and has received funding from the European Union's Horizon
2020 research and innovation programme under grant agreement number
633053. The views and opinions expressed herein do not necessarily
reflect those of the European Commission. This research was supported
in part by grant ENE2012-30832, Ministerio de Econom\'{\i}a y
Competitividad, Spain.

\section*{References}

\end{document}